\documentclass[twocolumn,showpacs,prl]{revtex4}

\ifx\pdfoutput\undefined
\usepackage{graphicx}
\else
\usepackage[pdftex]{graphicx}
\usepackage{epstopdf}
\fi

\usepackage[center]{subfigure}

\begin{document}

\title{Dynamics of precipitation pattern formation at geothermal hot springs}

\author{Nigel Goldenfeld, Pak Yuen Chan and John Veysey}
\affiliation{Department of Physics, University of Illinois at
Urbana-Champaign, Loomis Laboratory of Physics, 1110 West Green Street, Urbana, Illinois, 61801-3080.}

\begin{abstract}

We formulate and model the dynamics of spatial patterns arising during
the precipitation of calcium carbonate from a supersaturated shallow
water flow.  The model describes the formation of travertine deposits
at geothermal hot springs and rimstone dams of calcite in caves.  We
find explicit solutions for travertine domes at low flow rates,
identify the linear instabilities which generate dam and pond formation
on sloped substrates, and present simulations of statistical landscape
evolution.

\end{abstract}


\pacs{05.45.Ra, 87.23.–n, 47.54.-r, 89.75.Kd, 47.20.Hw, 47.15.gm, 47.55.np}
\maketitle

\begin{figure}
\includegraphics[width=0.50\columnwidth]{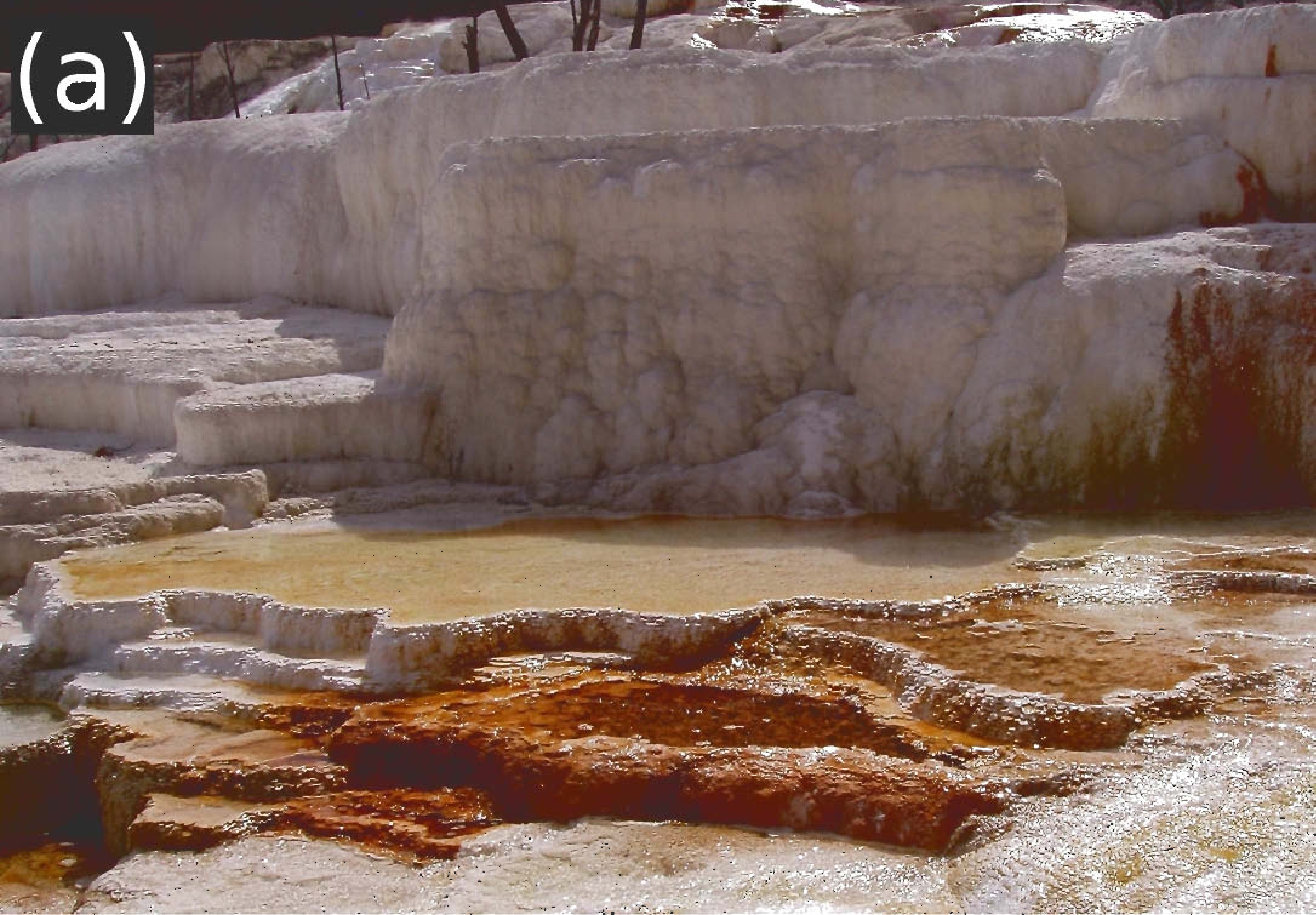}
\includegraphics[width=0.465\columnwidth]{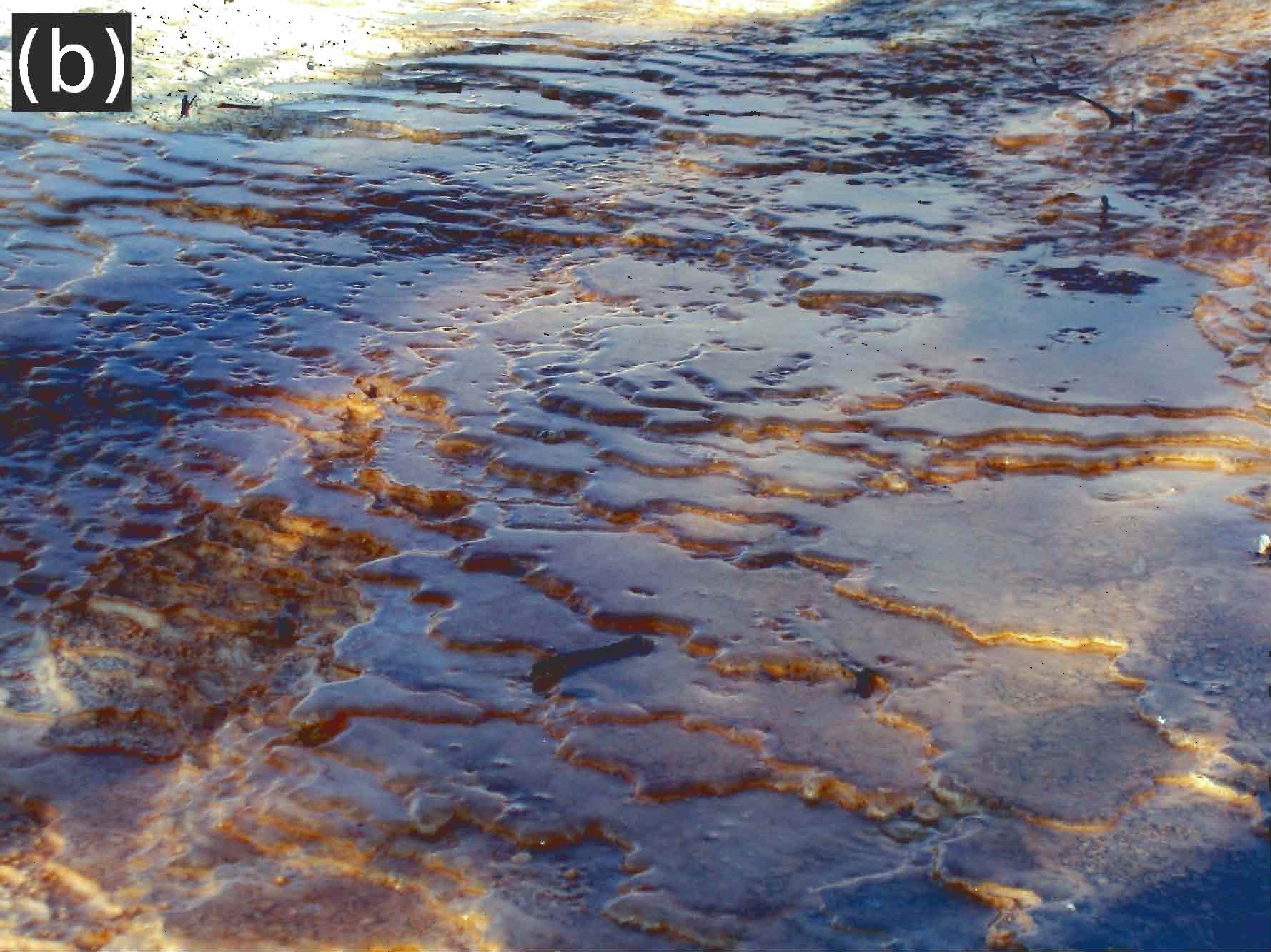}
\caption{(Color online) Travertine formation at Angel Terrace, Mammoth Hot Springs,
WY.  (a) a large pond, of order 1 meter in diameter, and smaller
features. (b) a portion of the flow system about 25 meters from
the vent, on the scale of centimeters.}\label{AT3}
\end{figure}

The terraced architecture of carbonate mineral deposits at geothermal
hot springs is one of the most striking and beautiful terrestrial
landscapes.  Spring water at above 70$^o$C emerges from the ground at a vent,
releases CO$_2$ and precipitates CaCO$_3$ in the form of travertine, as
it flows downhill over the pre-existing terrain\cite{ALLE35,
PENT94, FORD96}. The terrain itself is thus constantly changing in
response to the influx of CaCO$_3$, with measurements
indicating precipitation rates as high as 1-5 mm per day\cite{FRIE70,
PENT90, FOUK00}. The ever-changing substrate modifies the flow path of
the spring water, resulting in a constant dynamic interplay between the
landscape and the fluid flow as the travertine outcrop develops and
grows.  The resulting morphology is a cascade of nested ponds and
terraces at a wide variety of scales ranging from hundreds of meters
down to millimeters, as shown in Fig. (\ref{AT3}).

Most studies of this phenomenon have tended to focus on the microscopic
origins of the crystal growth process: what is the extent of
biomineralization due to thermophilic microbes\cite{FOUK00,
FOUK01}?; what are the primary controls on degassing\cite{HERM87,
ZHAN01}, mineral composition\cite{BARN71, CHAF91}, crystal
fabric and habit\cite{BUCZ91}, and structure\cite{BUSE86, RENA96}?
These issues are complex and system dependent, but a main theme is the
competition between biotic and abiotic mechanisms of precipitation.
Microbial metabolic activity can in principle locally influence the
CO$_2$ composition of the water column, thereby affecting the rate of
CaCO$_3$ precipitation. However, turbulent flow over pond lips
or evaporation can be more effective degassing mechanisms\cite{ZHAN01}.
In fact, the issue of biotic versus abiotic mechanism is not
straightforward, because even if the main mechanism for degassing is
abiotic, the kinetics of precipitation and crystal growth may require
the presence of exogenous particles (such as dead or living bacteria)
to initiate nucleation.

\begin{figure}
\includegraphics[width=0.47\columnwidth]{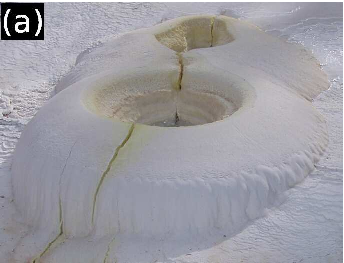}
\includegraphics[width=0.48\columnwidth]{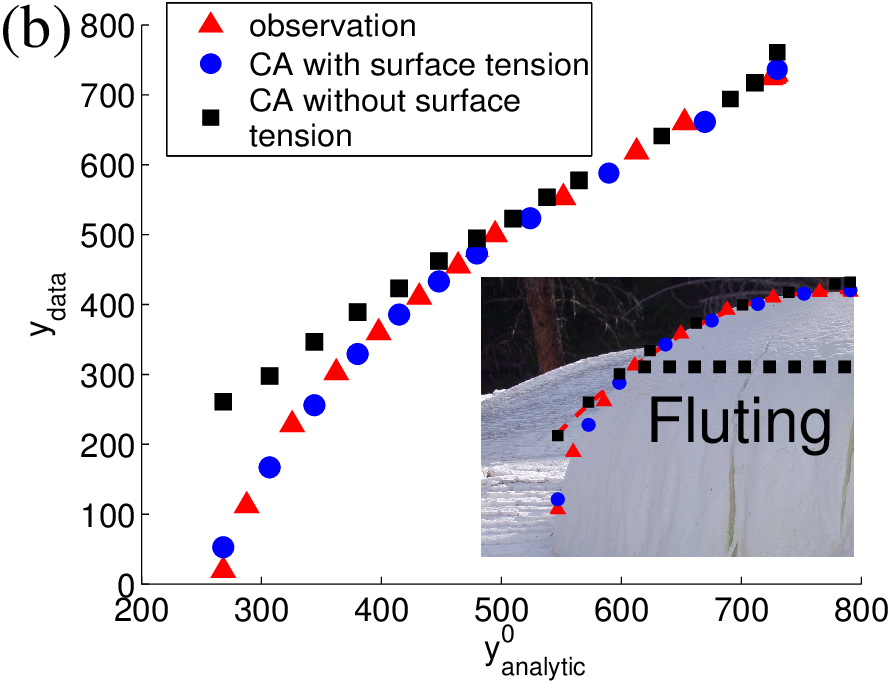}
\caption{(Color online) Travertine dome at Mammoth Hot Springs, WY. (a)
Dome whose central pond is 50.3cm in diameter. (b) Dome height as a
function of $\theta$.  Horizontal axis: analytical prediction from Eq.
(\ref{dome_shape}).  Vertical axis: simulated dome (CA) with no surface
tension (squares), with surface tension (circles), and the profile of
the dome in (a) (triangles, data digitized from photograph).  The
points all fall on a straight line with unit slope above the fluting
point, indicating agreement with Eq. (\ref{dome_shape}) fitting only
$r_0$. Below that one extra parameter is required, as explained in the
text.  Inset: comparison with dome from (a).}\label{Dome}
\end{figure}

The purpose of this Letter is to address instead the origin of the
large scale structure of the terraced architecture.  The ubiquity of
this morphology at carbonate geothermal hot springs world-wide, as well
as at low temperature speleothem rimstone dam formations\cite{MOTY05},
suggests that it is appropriate to seek a generic explanation based on
principles of fluid dynamics, precipitation kinetics, and crystal
growth dynamics, rather than one that hinges on specific material
parameters or system components such as microbes.  In fact, we will see
that it is possible to explain in this way the formation of ponds, the
variations in terrace morphology, the absence of an obvious
characteristic lengthscale of the landscape, and even the quantitative
properties of simple landscape motifs, such as the circularly-symmetric
travertine dome shown in Fig. (\ref{Dome}a).

The formation of a terraced architecture is a result of a depositional
instability arising from turbulent flow of a supersaturated solution
over a surface, and represents an example of free boundary dynamics in
precipitative pattern formation, related to the phenomena that give
rise to stalactites\cite{SHOR05a, SHOR05b}.

\smallskip
\noindent {\it Travertine domes:-\/} We begin by analyzing the dynamics
of pattern motifs, ignoring interactions between them, by analogy with
earlier work on solidification patterns\cite{BENJ83}.  In the
present problem, these features are pond lips\cite{WOOD91} and
travertine domes.

We now formulate a boundary-layer model\cite{BENJ83} of the growth of
travertine domes, coupling the evolution of the travertine substrate to
the fluid dynamics in a thin film of depth $h$ around it.  The
kinematic equation governing time evolution of the curvature, $\kappa$,
of a curve in two dimensions is given by\cite{BENJ83, BROW84}:
\begin{equation}
\frac{\partial\kappa}{\partial t}\bigg|_{\theta} = -\kappa^2\left(1+
\frac{\partial}{\partial \theta^2}\right)v_n,
\label{growth_eqn}
\end{equation}
where $\theta$ is the angle between the surface and the vertical axis
and $v_n$ is the normal growth velocity of the surface. The time
derivative in the equation is defined with respect to fixed $\theta$.
This equation is purely geometrical; for any given function $v_n$ of
water chemistry, surface kinetics, fluid flow state, the evolution of
$\kappa$ is determined. Here, we follow Wooding\cite{WOOD91} and make
the simple assumption that $v_n$ is directly proportional to the
depth-averaged tangential fluid velocity $U$: $v_n = G U$ where $G$ is
a mass transfer coefficient, depending on the water chemistry and the
turbulent flow near the growing surface\cite{CAMP82}.

In general, we need to couple the Navier-Stokes equation to Eq.
(\ref{growth_eqn}) to obtain a complete description of the coupled
fluid dynamics and surface kinematics, but there are a number of
simplifications.  First, because the growth rate is of order $1-5
\mathrm{mm/day}$ and the fluid flow rate is of order $1
\mathrm{mm/sec}$, there is a separation of time scales. So if we are
interested in the morphological evolution, we can neglect the change in
flow rate, represented by the time derivative in the Navier-Stokes
equation. Second, our field observations indicate that the thickness of
the fluid film flowing over the domes is very small compared to the
curvature of the surface; thus, we make the approximation that the
fluid is flowing down a (locally) constant slope.  In addition, the
flow is apparently laminar, so that we can use the Poiseuille-Hagen
profile for the velocity in thin film to give a depth-averaged mean
flow velocity $U$:
\begin{equation}
U = \left( \frac{\alpha\sin\theta}{r^2} \right)^{1/3},
\label{average_v}
\end{equation}
where $\alpha\equiv g Q^2/12\pi\nu$, $g$ is the gravitational
acceleration, $Q$ is the total mass flux coming out of the vent, $\nu$
is the viscosity of the fluid and $r$ is the axial distance from the
vent. Circular symmetry is imposed to arrive at (\ref{average_v}).  We
will later see that the assumption of laminar flow is self-consistently
verified.  For a dome, steadily translating upwards without change of
shape with velocity $v_t$, Eq. (\ref{growth_eqn}) gives
\begin{equation}
G\left(\frac{\alpha\sin\theta}{r^2}\right)^{1/3} = v_t\cos\theta,
\end{equation}
Rearranging terms gives the shape of the dome as a one-parameter family
of curves
\begin{equation}\label{dome_shape}
r(\theta)/r_0 = \sqrt{\frac{\sin\theta}{\cos^3\theta}}.
\end{equation}
where the scale factor $r_0\equiv \sqrt{G^3\alpha/v_t^3}$.  Eq.
(\ref{dome_shape}) is plotted in Fig. (\ref{Dome}b).  Good agreement is
obtained between our theory and the observations below a critical angle
$\theta_c$. From the fit, and the typical parameter values $G\sim
10^{-8}$, $v_t\sim 1\mathrm{mm/day}$ and $Q\sim
1\mathrm{cm}^3\mathrm{/sec}$, we obtain $U\sim 25 \mathrm{mm/sec}$ and
$h\sim1-10\mathrm{mm}$, and a Reynold's number, $\mathrm{Re}\equiv
Uh/\nu\sim 10-100$. The assumption of laminar flow is self-consistently
verified.

For angles $\theta > \theta_c$, the analytical profile does not fit
the field observations; the point of departure closely follows the
point where we also observe a fluting pattern around the dome.  We will
show below that this is due to the effects of surface tension at the
air-water-travertine interface. As the water flows out of the dome, it
is spread over an increasingly larger area and thus the fluid thickness
decreases, ultimately reaching a point where contact lines form and
surface tension can not be ignored.  The analytical solution above
neglects surface tension, but is able to predict the limiting point of
its validity when $h\sim d_c$, leading to a prediction for the scaling
dependence of the critical angle on the model parameters.

The inclusion of surface tension introduces an additional length scale,
namely, the capillary length, $d_c$, into the problem. Now, the only
other length scale in the problem is $r_0= \sqrt{gG^3Q^2/\nu v_t^3}$.
Since $\theta_c$ is dimensionless, it can only depend on the ratio
$r_0/d_c$ and $G$.  For a
given chemical environment, $G$ is fixed and we are left with the
prediction, derived from our analytical solution, that
\begin{equation}
\theta_c= \hat{f}(\sqrt{(gQ^2/\nu v_t^3)}/d_c),
\label{contact_line}
\end{equation}
verified below in Fig. (\ref{sim_terrace}b).

\smallskip
\noindent {\it The damming instability:-\/} To understand the lack of
any characteristic scale in the landscape, we consider the stability of
the moving boundary problem for turbulent fluid flowing down a constant
slope, on which deposition may occur.  In order to capture the
turbulent flow, we make two approximations.    First, we use the thin
film approximation attributed to de St. Venant, which is valid when the
fluid film thickness is much less than the characteristic scale of
variation of the flow in the streamwise direction, but include the
lowest order corrections for the curvature of the underlying
surface\cite{DRES78,SIVA81}.
\begin{equation}\label{Dressler_eqn}
\begin{array}{lll}
\partial_t u_0 + \partial_s E &=& {-C_fgu^2}/{gh(1-\kappa h)}\\
(1-\kappa h) \partial_t h -\partial_s q &=& 0\\
\end{array}
\end{equation}
where
\begin{equation}
\begin{array}{lll}
u(s,n,t)&=&{u_0(s,t)}/{(1-\kappa h)},\\
E(s,t)&=&\zeta+h\cos\theta+{u_0^2}/{(2g (1-\kappa h)^2)},\\
q(s,t)&=&{-u_0}/{(\kappa}\log(1-\kappa h)),
\end{array}
\label{Dressler_defs}
\end{equation}
where $u$ is the fluid velocity, $h$ is the fluid thickness, $\zeta$ is
the height of the underlying surface measured from a fixed horizontal
axis and $s$ is the arc length measured from the top of the system.
Secondly, we have modeled turbulent flow phenomenology by including the
term proportional to $C_f$, the Ch\'ezy coefficient\cite{CHEZ76}, which
empirically describes the energy lost due to turbulence, in a manner
consistent with Kolmogorov's 1941 scaling theory of turbulence
(K41)\cite{KOLM41,SREE99}.  These two equations have to be solved
together with the growth equation (\ref{growth_eqn}). The trivial
solution to this set of equations can be easily found, and is simply
uniform viscous flow down a slope.

To study the linear stability of this solution, we add a perturbation
proportional to $\exp(ikx+\omega (k) t)$, and calculate the spectrum of
the growth rate $\omega$ as a function of the wavenumber $k$ for the
linearised set of equations.  It is found that Re($\omega$) is positive
for all values of $k$, indicating that the system is unstable towards
perturbations of all length scales, with no special scale singled out
at linear order.

We next present an approach to simulate the statistical correlations
of the terraced landscape in the fully-nonlinear regime.

\smallskip
\noindent {\it Cellular modeling:-\/} We represent the landscape as
composed of stacked \lq\lq bricks" or cells by its height $H(i,j)$,
above a horizontal reference plane, where $i$ and $j$ are $x-y$
coordinates in the reference plane.  The water column is situated above
the height field and represented by the variable $W(i,j)$ describing
the volume of water above each coordinate element.  Each packet of
water can also contain calcium ions, $C(i,j)$,  and dissolved carbon
dioxide vapor, $V(i,j)$, which may potentially cause precipitation
through a caricature of the complex reaction pathway given by Ca$^{2+}$
+ 2HCO$_3^-$ $\rightleftharpoons$ CaCO$_3$(s) + H$_2$O + CO$_2$(g).  The
evolution of the landscape is governed by update rules on these fields,
which mimic and go beyond the continuum description given above.  In
addition, at each point, it is necessary to keep track of water that is
ponded, $W_p(i,j)$, and the temperature of the water $T(i,j)$.

A complete lattice update consists of the following steps, which we
will describe in more detail below: (1) Add water to the system, at the
spring source taken to be the origin; (2) Propagate all the water
in the system, by moving packets to nearest and next nearest neighbour
grid points, and ensuring that the water in all ponds is level; (3)
Update the water chemistry, e.g. $C(i,j)$ and $V(i,j)$ to take into
account outgassing due to fluid motion and depletion of Ca ions due to
precipitation; (4) Evolve the height field in response to the
precipitation of CaCO$_3$.

In step (1), a quantity of water $\delta W$ is added at the source,
so that $W^\prime (0) = W_0$, a
constant value appropriate for a constant pressure head (neglecting the
change in pressure due to the vertical growth of the landscape).  Here
and below, primed quantities denote the updated variables.  The new
water added to the system contains initial concentrations of calcium,
$C_0$, carbon dioxide, $V_0$, and is at a temperature $T_0$; the fact
that the source water is undersaturated is represented by $C_0<V_0$.
The values of these fields at the source point are updated based on the
volumetric ratio of the amount of water added to the existing water:
e.g. $C^\prime = (C(0)W(0) + \delta W C_0)/W_0$, and similarly for the
CO$_2$ concentration and temperature fields.

The transport of water in step (2) is carried out by a variation of an
algorithm used for braided river flow\cite{MURR94}, in which the flux
along bonds connecting a given lattice point to one of its eight
closest neighbours is determined by the landscape gradient along that
direction, while conserving the total volume of water.  If the slope
$S$ exceeds a threshold $S_c$, the flow is taken to be turbulent and
the flux is proportional to $\sqrt{S}$ in accord with Ch\'ezy's law.
The appropriate height variable for determining the chemical potential
of the water, and hence the equilibrium filling of ponds in the
landscape is the \lq\lq energy surface" $H_T \equiv H + W$.  Water
moving on this surface is also subject to surface tension and contact line
effects, that can be important near the rim of a pond, for
example\cite{KERR96}. This is modeled by requiring that water is
propagated only if $W$ exceeds a small threshold at that point.
Packets of water carry with them advected variables $T$, $C$ and $V$,
which are updated by the volumetrically-weighted average of all the
neighbourhood points from which the water packet originated.

Water chemistry is updated in step (3), by allowing a constant fraction
of CO$_2$ to outgas at each time step, with an additional outgassing
component proportional to $\sqrt{S}/(1+\sqrt{S})$ to reflect the
influence of slope-initiated turbulent flow.  As CaCO$_3$ is deposited,
the concentrations $C$ and $V$ change to reflect mass balance.

The evolution of the height field $H$ is the final step in the lattice
update, with a change given by $\delta H=(C-V)\times (R_1 +
R_2\vec{F}\cdot\vec{n} + R_3(S))$.  Here $R_1$ and $R_2$ are positive
constants, $\vec{F}$ is the flux between cells, with a direction given
by the component of the gradient of the energy surface $H_T$ between
cells and magnitude given by the volume of water propagation per unit
time step, $\vec{n}$ is the unit vector normal to the underlying
surface $H$, and $R_3(S)$ is proportional to $\sqrt{S}$, representing
the increased precipitation due to Bernoulli effects and local
turbulent degassing.

\begin{figure}
\includegraphics[width=0.39\columnwidth]{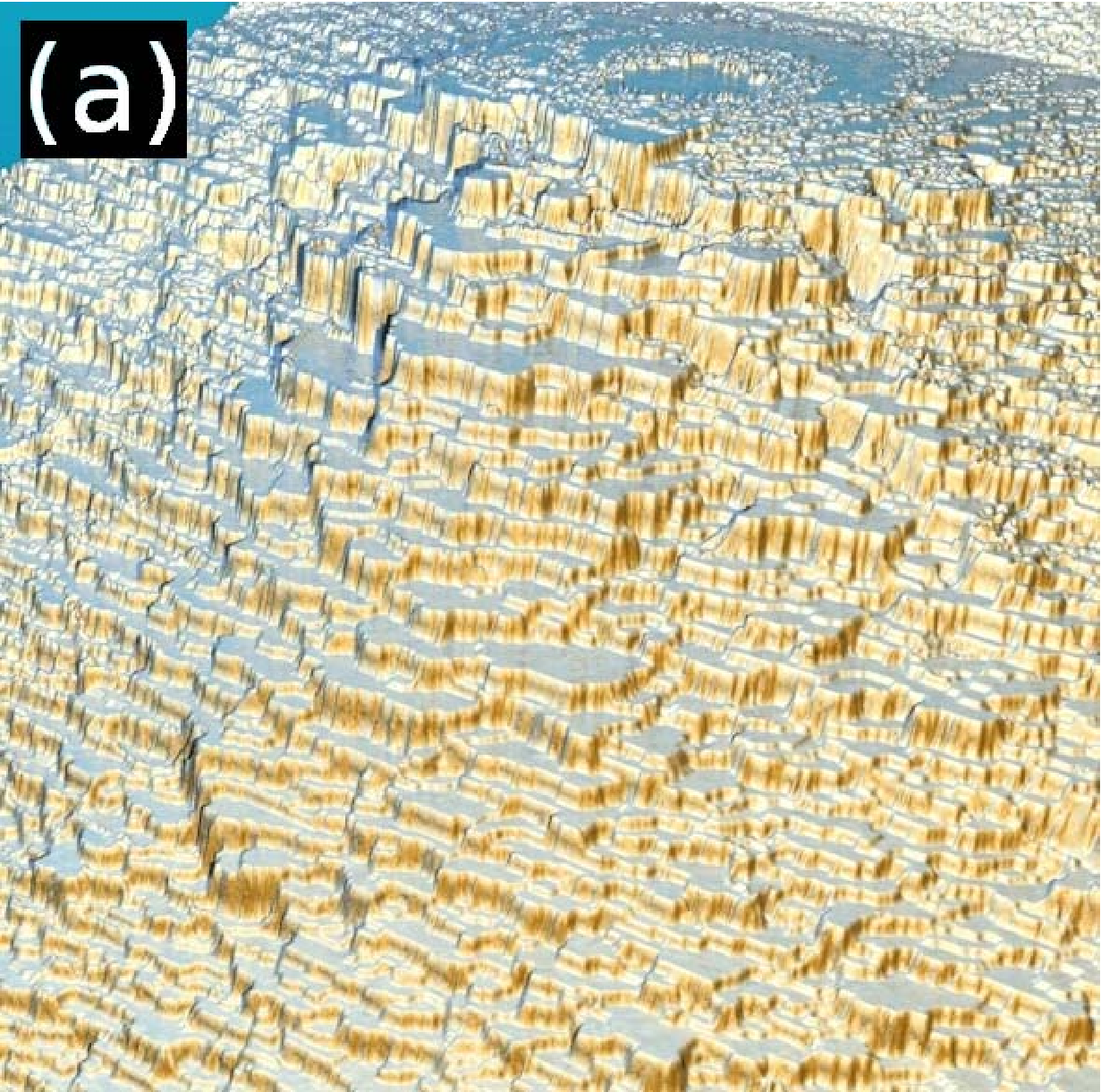}
\includegraphics[width=0.55\columnwidth]{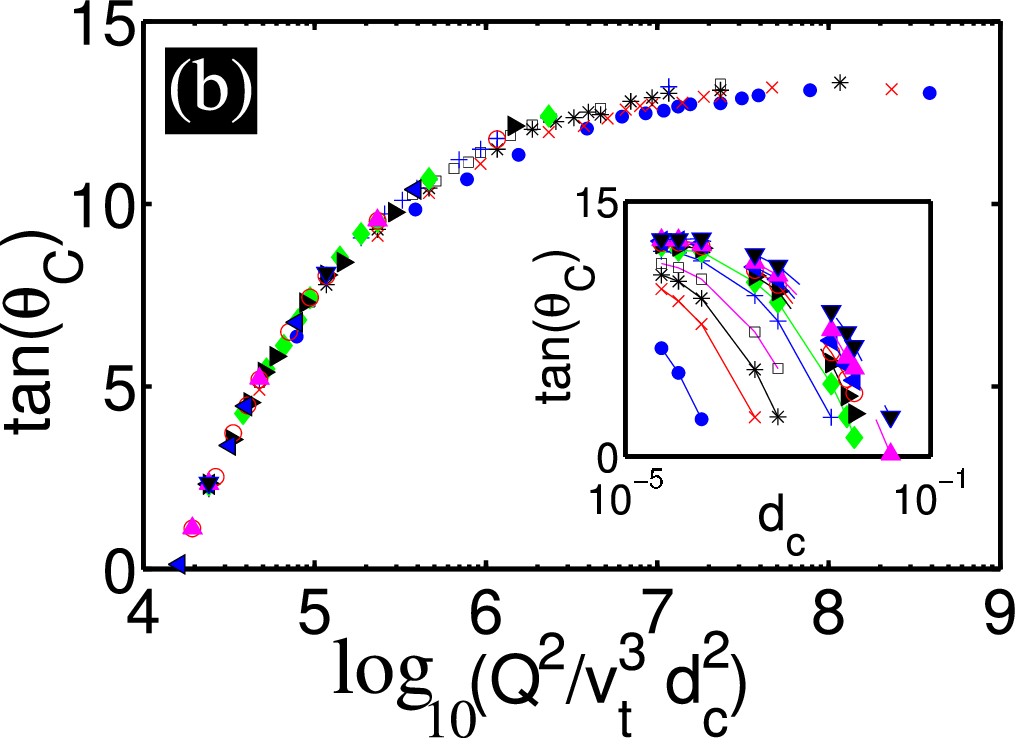}
\caption{(Color online) Results from simulation of the CDS model. (a)
Portion of a typical simulated landscape. The figure clearly shows
distinct geomorphological regimes or facies, as observed in the
field\cite{FOUK00}. (b) Critical angle for the contact line formation
on a travertine dome, plotted according to Eq. (\ref{contact_line}),
showing data collapse, as predicted by theory.  Inset: raw
data.}\label{sim_terrace}
\end{figure}

In Fig. (\ref{sim_terrace}a) is shown a snapshot from a typical
time-dependent simulation, initiated on a sloping plane with small
initial roughness.  For generic values of the model parameters, we
observe the depositional instability predicted above, and the formation
of ponds and terraces in broad qualitative agreement with field
observations. The same calculation also yields travertine domes when
initiated on an initially horizontal surface. Fig. (\ref{Dome}b) shows
that the CA model agrees with the analytical theory when the surface
tension is switched off. Moreover, the same CA model predicts the exact
shape of the observed travertine domes when surface tension is switched
on, with a contact line at $\theta_c$ where fluting emerges.  The
scaling prediction Eq. (\ref{contact_line}) is verified in Fig.
(\ref{sim_terrace}b) over 5 decades of $Q^2v_t^3/d_c^2$; the inset shows
the raw (unscaled) data for $\theta_c$ as a function of $d_c$ for $v_t
\sim 0.5$ and $0.1 < Q < 5000$.  Our results show that travertine
precipitation pattern formation results from an interplay between fluid
flow, capillarity and chemistry, that can be captured quantitatively
from both analytical and cell dynamical system approaches.

Future field work and extended simulation studies will explore the
statistical properties of terraced landscapes, and whether these have a
universal character computable with no adjustable parameters by minimal
models. For example, layer fluctuations in ancient stromatolite rocks
exhibit power-law correlations accurately captured by a simple generic
model\cite{GROT96} in the same spirit as the one given here.  Our
preliminary simulations indicate that at least some statistical
measures of the landscapes, such as the pond size distribution
function, are fractal.

We acknowledge valuable discussions with Bruce Fouke and Michael
Kandianis.  We thank Nicholas Guttenberg for assistance with image
rendering.  This work was supported in part by the National Science
Foundation through grant number NSF-EAR-02-21743.

\bibliographystyle{apsrev}

\bibliography{travertine}

\end{document}